\documentstyle[epsf,12pt]{article}
\def\beeq{\begin{eqnarray}} \def\eeeq{\end{eqnarray}}
\newcommand\mysection{\setcounter{equation}{0}\section}
\renewcommand{\theequation}{\thesection.\arabic{equation}}
\newcounter{hran} \renewcommand{\thehran}{\thesection.\arabic{hran}}

\def\bmini{\setcounter{hran}{\value{equation}}
  \refstepcounter{hran}\setcounter{equation}{0}
  \renewcommand{\theequation}{\thehran\alph{equation}}\begin{eqnarray}}

\def\bminiG#1{\setcounter{hran}{\value{equation}}
\refstepcounter{hran}\setcounter{equation}{-1}
\renewcommand{\theequation}{\thehran\alph{equation}}
\refstepcounter{equation}\label{#1}\begin{eqnarray}}

\def\emini{\end{eqnarray}\relax\setcounter{equation}{\value{hran}}\renewcommand{\theequation}{\thesection.\arabic{equation}}}

\def\ben{\begin{enumerate}}  \def\een{\end{enumerate}}
\def\bit{\begin{itemize}}    \def\eit{\end{itemize}}
\def\beq{\begin{equation}}   \def\eeq{\end{equation}}
\def\bea{\begin{eqnarray}}  \def\eea{\end{eqnarray}}
\def\nn{\nonumber}
\def\noi{\noindent}

\def\lsim{\raise0.3ex\hbox{$<$\kern-0.75em\raise-1.1ex\hbox{$\sim$}}}
\def\gsim{\raise0.3ex\hbox{$>$\kern-0.75em\raise-1.1ex\hbox{$\sim$}}}
 \def\cite#1{[\ref{#1}]}
 \def\citd#1#2{[\ref{#1},\ref{#2}]}
 \def\citt#1#2#3{[\ref{#1},\ref{#2},\ref{#3}]}
 \def\citm#1#2{[\ref{#1}--\ref{#2}]}
 
\def\cits#1#2#3#4{[\ref{#1}--\ref{#2},\ref{#3},\ref{#4}]}
\def\citn#1#2#3#4{[\ref{#1},\ref{#2},\ref{#3},\ref{#4}]}

\begin{document}
\vbox to 1 truecm {}
\begin{center}
{\bf CONSTRAINTS FROM CHARGE AND COLOUR BREAKING MINIMA IN THE (M+1)SSM 
} \\
\vspace{1 truecm}
{\bf Ulrich Ellwanger and Cyril Hugonie}
\footnote{email: ellwange@qcd.th.u-psud.fr,
cyrilh@qcd.th.u-psud.fr}\\ Laboratoire de Physique Th\'eorique
\footnote{Unit\'e mixte de Recherche (UMR 8627)}\\    
Universit\'e de Paris XI, Centre d'Orsay, B\^atiment 210, 91405
Orsay Cedex, France\\  
\end{center}
\vspace{2 truecm}
\begin{abstract}
We study the constraints on the parameter space of the supersymmetric
standard model extended by a gauge singlet, which arise from the absence
of global minima of the effective potential with slepton or squark
vevs. Par\-ti\-cu\-lar attention is paid to the so-called ``UFB''
directions in field space, which are $F$-flat in the MSSM. Although
these directions are no longer $F$-flat in the (M+1)SSM, we show that
the corresponding MSSM-like constraints on $m_0/M_{1/2}$ apply also to
the (M+1)SSM. The net effect of all constraints on the parameter space
are more dramatic than in the MSSM. We discuss the phenomenological
implications of these constraints.  \end{abstract} 

\vspace{2 truecm} 

\noi LPTHE Orsay 98-81\\ 
\noi January 1999 \\
 
\newpage
\pagestyle{plain}
\mysection{Introduction}
\hspace*{\parindent}
In any supersymmetric extension of the standard model the vevs
of squarks and charged sleptons have to vanish in order not to break the
gauge symmetries $SU(3)_c$ and $U(1)_{em}$ spontaneously. In the last
years, many investigations of the MSSM have been carried out in order to
find the constraints on the parameter space implied by the absence of
such vevs \citm{1r}{17r}. The aim of the present paper is to find
corresponding constraints in the (M+1)SSM, the supersymmetric extension
of the standard model with an additional gauge singlet superfield $S$ in
order to replace the $\mu$ term in the superpotential by a vev $\langle 
S \rangle$ [18,1,4,10,19]. (Often the (M+1)SSM is also referred to as the 
NMSSM, the next-to-minimal supersymmetric standard model). 
\par

First, we briefly review the dangerous directions in field space
and the cons\-traints on the parameter space in the MSSM. Then, in
section 2, we compare the most relevant constraints in the MSSM to
the ones in the (M+1)SSM. \par 

Let us, to start with, recall the general structure of the scalar potential
$V(\varphi_i)$ of any supersymmetric extension of the standard model:
1) positive semi-definite $F$- and $D$-terms; 2) soft susy breaking
mass terms of the form $m_i^2|\varphi_i|^2$ (except for the $B$-term
$\mu B H_1 H_2 + h.c.$ in the MSSM); 3) soft susy breaking
trilinear couplings of the form $A_{ijk} \varphi_i \varphi_j 
\varphi_k$; 4) radiative corrections to the effective potential. \par 

These radiative corrections can actually be made to vanish, if an appropriate
renormalisation scheme and an appropriate renormalisation scale $Q$
(with $Q \sim \ \langle \varphi \rangle $) for all parameters appearing in
$V(\varphi_i)$ are chosen \cite{20r}. In practice, where minimal
subtraction schemes are employed, the appropriate choice of $Q$ cancels
the radiative corrections only for $\langle \varphi \rangle \ 
\gg m_i,A_{ijk}$, and one is left with the so-called Coleman-Weinberg 
contributions in the regime $\langle \varphi \rangle \ \sim m_i, A_{ijk}$ 
\citt{20r}{9r}{21r}. \par

Now, the dangerous directions in field space and the corresponding
constraints can be classified as follows: \par

1) Traditional CCB bounds \cits{1r}{4r}{9r}{10r}: here negative
contributions to the scalar potential arise from one of the trilinear
couplings $A_{ijk} \varphi_i \varphi_j \varphi_k$ (and appropriate
phases of the fields $\varphi_i$). The $D$-terms can be made to vanish
by choosing $\langle \varphi_i \rangle =\langle \varphi_j \rangle = 
%\break \noindent 
\langle \varphi_k \rangle $, but some of the $F$-terms are always
non-zero. Assuming, at the GUT scale, universal trilinear couplings
$A_0$, scalar masses $m_0$ and gaugino masses $M_{1/2}$, the absence of
such minima leads to upper limits on $A_0$ as a function of $m_0$ and
$M_{1/2}$ \cits{2r}{4r}{9r}{10r}. \par 

2) So-called UFB bounds [8,12-14,16,17]: these arise from directions in
field space which are both $D$-flat and $F$-flat. Then, the
contributions from the trilinear couplings vanish as well. Dangerous
directions thus involve fields $\varphi_i$, where the soft susy breaking
mass $m_i^2$ is negative (at least at small scales). Since, in order to
trigger $SU(2) \times U(1)_Y$ symmetry breaking, the mass $m_1^2$ of the 
Higgs scalar $H_1$ (which couples to the top quark in our convention) is
typically negative, one of the fields $\varphi_i$ is always given by
$H_1$. With mass parameters $m_1^2$ and some $m_i^2$ taken at the scale
$Q = M_Z$, one can then find directions $\{H_1, \varphi_i\}$ in field
space, with respect to which the scalar potential is possibly
unstable. If one neglects the scale dependence of $m_1^2$ and $m_i^2$,
one arrives even at the conclusion that the scalar potential is
unbounded from below in these directions (therefore the notion
``UFB''). Clearly this conclusion turns out to be erroneous, once the
scale dependence of the masses is correctly taken into account, and if
the masses squared are assumed to be positive at some large scale
$M_{GUT}$. \par 

Nevertheless constraints on the parameters arise from the absence of
true minima of the scalar potential in such directions, and these
constraints are still called ``UFB bounds''. Typically, assuming
universal soft terms, one obtains lower limits on the ratio
$m_0/M_{1/2}$ of ${\cal O}(1)$ (depending to some extent on other
parameters like $\tan \beta$, $h_t$ or $M_{1/2}$ alone). \par

3) Improved CCB bounds \citn{6r}{7r}{11r}{12r}: It has been observed
that the directions in field space 1) and 2) above (with vanishing
$D$-terms) do not necessarily allow to find the absolute minimum of the
scalar potential. Allowing for more complicated combinations of vevs
$\varphi_i$, and some $D$-terms to be non-zero (implying, typically,
$\varphi_i \not= \varphi_j$), deeper minima can often be
obtained. Sometimes these directions in field space interpolate between
the directions 1) and 2) above. Usually, the corresponding constraints
depend in a complicated way on many parameters of the MSSM (the soft
terms, $\mu$, and the Yukawa couplings) and cannot be represented in the
form of universal inequalities among just two or three parameters. \par

The actual relevance of the bounds 1) -- 3) above is not entirely
evident: Even if a charge and/or colour breaking minimum of the scalar
potential exists, which is deeper than the standard $SU(2) \times
U(1)_Y$ breaking minimum, this situation can be acceptable if the
tunneling rate out of the standard minimum is small compared to the age
of the universe. Only the cases with large tunneling rates can
definitely be excluded. These tunneling rates have been estimated in
many papers [3,5,13,15,16] with the result that they are often quite
small. Then, the relevance of the bounds 
\break \noindent 1) -- 3) above depends on the
early cosmology, i.e. into which minimum we drop after 
inflation. Since the answer depends on the inflationary potential and
the reheating temperature \cite{22r}, this question cannot be resolved
in terms of the parameters of the MSSM above. \par

In the following we will study the constraints arising from lower lying
charge and colour breaking minima of the scalar potential leaving the
question of tunneling rates aside. In the next chapter we discuss some
particularly relevant directions in field space in some detail, in order
to compare the corresponding constraints on the parameters of the MSSM
with the (M+1)SSM.  

\mysection{Constraints in the MSSM and (M+1)SSM}

\hspace*{\parindent} First, we consider the most dangerous CCB direction
of type 1), which involves the trilinear coupling $h_e A_E E_{R,1} L_1 \-
\cdot \- H_2$. Here $E_{R,1}$ is the right-handed selectron, $L_1$ the
left-handed slepton doublet of the first generation, and $H_2$ the
corresponding Higgs doublet. $h_e$ denotes the electron Yukawa coupling
with $h_e \sim 10^{-5}$. From the absence of a non-trivial minimum of
the scalar potential in the $D$-flat direction $|E_{R,1}| = |L_1| =
|H_2|$ the following inequality among the soft susy breaking terms can
be derived in the MSSM \citn{2r}{4r}{9r}{10r}:    

\beq
\label{2.1}
A_E^2 < 3 \left ( m_E^2 + m_L^2 + \widehat{m}_2^2 \right ) 
\eeq    

\noi with $\widehat{m}_2^2 = m_2^2 + \mu^2$, and where $m_E^2$, $m_L^2$
and $m_2^2$ are the soft susy breaking mass terms associated with the
three fields above. If the inequality (\ref{2.1}) is violated, the
fields develop vevs of ${\cal O}\left ( A_E/h_e \right )$, and the depth
of the minimum is of ${\cal O}\left ( A_E^4/h_e^2 \right )$. Accordingly
the inequality (\ref{2.1}) has to be imposed at a scale $Q \sim A_E/h_e
\sim 10^7$~GeV. \par 

In the (M+1)SSM, there is no $\mu$ term; an effective $\mu$ term is
generated once the vev $\langle S \rangle$ is non-zero. However, once the 
inequality
(\ref{2.1}) is violated with $\mu = 0$, the minimum in the corresponding
CCB direction is deeper than the minimum associated with a non-zero vev
$\langle S \rangle $ (since $h_e$ is extremely small). Accordingly the 
inequality
(\ref{2.1}) holds in the (M+1)SSM with $\widehat{m}_2^2 =
m_2^2$. Assuming universal soft terms at the GUT scale, the inequality
(\ref{2.1}) then  becomes \cite{10r}

\beq
\label{2.2}
\left ( A_0 - 0.5 \ M_{1/2} \right )^2 < 9 \ m_0^2 + 2.67 \ M_{1/2}^2
\quad . 
\eeq

\noi In the MSSM, the inequality (\ref{2.2}) is weakened by additional
positive terms involving $\mu_0^2$ on the right-hand side. \par

Also, CCB minima in the stop direction $|T_R| = |Q_3| = |H_1|$ can be
considered \citt{2r}{4r}{9r}. In the MSSM, the inequality (\ref{2.1})
has then to be replaced by  

\beq
\label{2.3}
A_t^2 < 3 \left ( m_{T_R}^2 + m_{Q_3}^2 + \widehat{m}_1^2 \right ) 
\eeq

\noi with $\widehat{m}_1^2 = m_1^2 + \mu^2$. If the inequality
(\ref{2.3}) is violated, the fields develop vevs of ${\cal O}\left
( A_t/h_t \right ) \sim {\cal O}(M_Z)$ (accordingly it has to be imposed
at the weak scale), and the depth of the minimum is of ${\cal O} \left
( A_t^4/h_t^2 \right )$. It is not evident, which of the two
inequalities (\ref{2.1}) and (\ref{2.3}) is more relevant in the MSSM:
the soft masses $m_{T_R}^2$, $m_{Q_3}^2$ and $m_1^2$ on the right-hand
side of (\ref{2.3}) can be small or even negative at the weak scale; on
the other hand, $A_t^2$ at the weak scale is also usually smaller than
$A_E^2$ at a scale $A_E/h_e$. Hence a case-by-case analysis is
required. \par 

The situation is somewhat simpler in the (M+1)SSM: The minimum
associated with a non-zero vev $ \langle S \rangle $ is always deeper 
(of ${\cal O}\left
( A_k^4/k^2 \right )$, see below) than a minimum in the stop
direction. Accordingly, if one compares minima in the stop direction with the
standard minimum, a non-zero vev $ \langle S \rangle $ has to be taken into
account. Thus the inequality (\ref{2.3}) holds in the (M+1)SSM
with an effective $\mu$-term included, in 
contrast to the inequality (\ref{2.1}). We found that, after imposing
(\ref{2.2}) and the present phe\-no\-me\-no\-lo\-gi\-cal constraints on
the parameters on the (M+1)SSM, the inequality (\ref{2.3}) with an 
effective $\mu$-term included is always satisfied automatically.
\par

Improved CCB bounds can be obtained from the absence of vevs in more general 
directions in field space with non-zero vevs of $T_R$, $Q_3$, $H_1$, $H_2$ 
and sleptons \citd{11r}{12r}. The corresponding constraints cannot be
represented in terms of simple inequalities among the bare parameters. 
However, since $m_{T_R}^2$ can be negative at the weak scale, 
non-trivial constraints follow already from the absence of a vev $\langle
T_R \rangle$ alone. Approximate analytic expressions for $m_{T_R}^2$ can be 
found, e.g., in \citn{10r}{16r}{17r}{23r}. One obtains, from these references, 
$m_{T_R}^2$ as a function of $M_{1/2}$, $m_0$, $A_0$ and 
$\rho = h_t^2/h_{t,QFP}^2 \ $: 

\beq
\label{2.4}
 m_{T_R}^2 \simeq (1-\rho) m_0^2 - {{\rho (1-\rho)} \over 3} 
(2.24 M_{1/2} - A_0)^2 + (6.6 - 2.6 \rho) M_{1/2}^2 \quad . 
\eeq

If the right hand side of eq. (\ref{2.4}) is negative, the depth of the 
minimum with $\langle T_R \rangle \neq 0$ is $\sim -{3 m_{T_R}^4 / 2 
g_3^2}$. (We have checked that this minimum is deeper than the one with 
$\langle T_R \rangle = \langle T_L \rangle \neq 0$, which would cancel the
$SU(3)$ $D$-term.) If the soft susy breaking terms and hence $|m_{T_R}^2|$ are
large, and/or if eq. (\ref{2.4}) is strongly violated, this stop minimum is
deeper than the standard one. We have performed a numerical scan of the complete
parameter space of the model, whose details are described in refs. [10,19]. 
For each point in the parameter space, which satisfies the inequality
(\ref{2.2}) and the present phenomenological constraints, we have compared the
depth of the potential in the stop direction to the standard minimum (with
radiative corrections due to stop/top loops included in both cases). The
corresponding constraints on the parameter space are discussed together with 
our other results below. \par

Now we turn to the ``UFB'' directions of type 2), which we will discuss
in some detail. Let us recall, to this end, the superpotential of the
MSSM: 

\bea
\label{2.5}
g_{MSSM} &=& \sum_i h_{u,i} \ Q_i \cdot H_1 \ U_{iR}^c + \sum_i h_{d,i}
\ Q_i \cdot H_2 \ D_{iR}^c \nn \\
&&+ \sum_i h_{\ell, i} \ L_i \cdot H_2 \ E_{ik}^c + \mu \ H_1\cdot H_2
\quad . 
 \eea   

\noi A particularly dangerous $D$- and $F$-flat direction in field space
has been identified by Komatsu \cite{8r}. It is associated with vevs of
the neutral component of $H_1^0$, the down squarks of the third
generation $D_{3R}^c$ and $D_{3L}$, and the stau $\widetilde{\nu}_3$:

\bea
\label{2.6}
&& \quad \langle H_1^0 \rangle  = H_1 \quad , \nn \\
&& \langle D_{3R}^c \rangle  = \langle D_{3L} \rangle  = d \quad , \nn \\
&& \qquad \langle \widetilde{\nu}_3 \rangle  = \widetilde{\nu} \quad .
 \eea

\noi For arbitrary $H_1$, $d$ and $\widetilde{\nu}$, the $U(1)_Y$
$D$-term and the third component of the SU(2) $D$-term are,
respectively, 

\beq
\label{2.7}
D_Y = {g_1 \over 2} \left ( d^2 + H_1^2 - \widetilde{\nu}^2 \right ) \ ,
\quad D_{SU(2)}^3 = - {g^2 \over 2} \left ( d^2 + H_1^2 -
\widetilde{\nu}^2 \right ) \quad .  \eeq 

\noi From the superpotential (\ref{2.5}) one finds that the only
$F$-term, which is a priori non-zero, is

\beq
\label{2.8} 
F_{H_2^0} = - h_{d,3} \ d^2 - \mu \ H_1 \quad .
\eeq    

\noi Hence all $D$- and $F$-terms vanish for

\bea
\label{2.9}
&&H_1 = - {h_{d,3} \over \mu} \ d^2 \quad , \nn \\
&&\widetilde{\nu}^2 = d^2 \left ( 1 + {h_{d,3}^2 \over \mu^2} \ d^2
\right ) \quad . 
\eea

\noi Then, the only non-vanishing terms in the scalar potential are mass
terms, and the potential along this direction becomes 

\beq
\label{2.10}
V(d) = \alpha d^4 + \beta d^2 
\eeq  

\noi with

\beq
\label{2.11}
\alpha = {h_{d,3}^2 \over \mu^2} \ \left ( m_1^2 + m_{L_3}^2 \right )
\quad , \qquad 
\beta = m_{Q_3}^2 + m_{D_3}^2 + m_{L_3}^2 \quad .
  \eeq  
  
At low scales, $m_1^2$ is usually negative, and $m_{L_3}^2$ is the
smallest susy breaking mass among the three left-handed slepton
doublets. Hence $\alpha$ can well be negative, and $V(d)$ seems to be
unbounded from below in this case. However, the appropriate scale
dependence of all parameters in $V(d)$ has to be taken into
account. Here, this appropriate scale $Q$ is 

\beq
\label{2.12}
Q \sim h_t|H_1| = h_t \ h_{d,3} \ d^2/\mu \quad , 
\eeq  

\noi and $V(d)$ should be written as

\beq
\label{2.13}
V(d) = \alpha \left ( Q^2(d) \right ) d^4 + \beta \left ( Q^2(d)
\right ) d^2 \quad 
.
  \eeq

\noi Thus, if $\alpha \left ( Q^2 = M_Z^2 \right )$ is negative, but all 
masses squared and hence $\alpha$ are positive at some large scale, $V(d)$ 
has a true minimum. \par 

The same reasoning applies to other $D$- and $F$-flat directions in
field space, as the one considered by Casas et al. in \cite{12r}, where
the down squarks are replaced by sleptons. \par

Recently, analytic approximations to the potential in these directions
have been studied by Abel and Savoy \cite{16r}, and conditions for
non-trivial minima as well as the corresponding tunneling rates have
been discussed. Actually it has been found that, even if a deeper
minimum in such a ``UFB'' direction exists, the decay rate of the
standard vacuum is usually negligible compared to the age of the
universe. Nevertheless, the condition for such a minimum not to be
deeper than the standard one implies a lower limit on the ratio
$m_0/M_{1/2}$ of ${\cal O}(1)$ [8,12-14,16,17] (assuming, again,
universal soft terms); from ref. \cite{16r} one finds, as a function of
$h_t$ or $\tan \beta$,

\beq
\label{2.14}
{m_0 \over M_{1/2}} > 0.3 - 1.0 \quad ,
\eeq  

\noi where the lower bound 0.3 corresponds to larger values of $\tan
\beta$. If the inequality (\ref{2.14}) is violated, one has to assume
that the early cosmology places one into the (local and metastable)
standard minimum of the potential. \par 

 All previous discussions of UFB directions concerned only the MSSM,
hence we turn now to the (M+1)SSM. It involves an additional singlet
superfield $S$, and the superpotential reads 

\bea
\label{2.15}
g_{(M+1)SSM} &=& \sum_i h_{u,i} \ Q_i \cdot H_1 \ U_{iR}^c + \sum_i
h_{d,i} \ Q_i \cdot H_2 \ D_{iR}^c \nn \\
&&+ \sum_i h_{\ell, i} \ L_i \cdot H_2 \ E_{ik}^c + \lambda S \ H_1
\cdot H_2 + {k \over 3} \ S^3 \quad .
 \eea

Let us have a look at the same direction (\ref{2.6}) in field space and
add, in addition, an arbitrary vev $s$ of the singlet scalar. The
$D$-terms are still given by eq. (\ref{2.7}), but now two $F$-terms are
a priori non-zero: 

\bea
\label{2.16}
&&F_{H_2^0} = - h_{d,3} \ d^2 - \lambda s \ H_1 \quad , \nn \\
&&F_S = k \ s^2 \quad .
\eea

One easily finds that both $F$-terms vanish only for $s = d= 0$. Hence
directions of the form (\ref{2.6}) in field space can no longer be
$F$-flat in the (M+1)SSM, and it seems that no ``UFB''-bounds exist in
this model. We will now show that this conclusion is wrong. \par

First, we assume that the Yukawa couplings $\lambda$ and $k$ in the
superpotential (\ref{2.15}) are small ($\lambda$, $k \ \lsim \
10^{-2}$), as it is the case in most of the parameter space of the model
\citt{10r}{18r}{19r}. Assuming, in addition, vevs of $H_1$ and $H_2$ of
${\cal O}(M_Z)$, the vev of $s$ is determined to a high precision
by the terms $V_s(s)$ in the scalar potential which depend solely on
$s$: 
   
\beq
\label{2.17}
V_s(s) = k^2 s^4 + {2 \over 3} k \ A_k \ s^3 + m_s^2 \ s^2 \quad . 
\eeq  

\noi The global minimum of $V_s$ is assumed at

\beq
\label{2.18} 
\bar{s} = -{A_k \over 4k} \left ( 1 + \sqrt{1 - {8m_s^2 \over A_k^2}}
\right ) 
\eeq 

\noi provided the parameters $A_k$, $m_s$ satisfy

\beq
\label{2.19}
A_k^2 \geq 9 \ m_s^2 \quad . 
\eeq

\noi If one plugs the vev (\ref{2.18}) of $s$ back into the complete
scalar potential of the \break \noindent (M+1)SSM, one obtains

\beq
\label{2.20}
V_{(M+1)SSM}(\varphi_i) = \bar{V}_{MSSM}(\varphi_i) + \lambda ^2 \left
( H_1 \cdot H_2 \right )^2 + V_s(\bar{s}) \quad .  
\eeq

\noi Here $\bar{V}_{MSSM}$ denotes the scalar potential of the MSSM with
effective $\mu$ and $B$ terms given by

\bea
\label{2.21}
&&\bar{\mu} = \lambda \bar{s} \quad , \nn \\
&&\bar{B} = A_{\lambda} + k\bar{s} \quad .
 \eea

Let us now consider the potential in "UFB" directions like (\ref{2.6}). With
$H_2 = 0$, the potential can be written as

\beq
\label{2.22}
V_{(M+1)SSM} (H_1, d, \widetilde{\nu},s) = \bar{V}_{MSSM} (H_1, d,
\widetilde{\nu},s) + V_s({s}) \quad ,
 \eeq

\noi where $V_{(M+1)SSM}$ depends on $s$ through the effective $\mu$ term given
in (\ref{2.21}). In principle, $V_{(M+1)SSM}$ should be minimized with respect
to $H_1$, $d$, $\widetilde{\nu}$ and $s$, and the depth of the corresponding
minimum should be compared to the depth of the "physical" minimum with $d = 
\widetilde{\nu} = 0$ (but $H_1$, $H_2$ and $s \neq 0$). Instead, we will now
compare the depths of the effective potential (with radiative corrections
included) at three different points in field space: \par

\noi Point A: the "physical" vacuum with $\langle s \rangle = s_{phys}$ (which
is possibly, but not necessarily close to $\bar s$ of eq. (\ref{2.18})). \par

\noi Point B: the minimum of $V_{(M+1)SSM} (H_1, d, \widetilde{\nu},s_{phys})$
with respect to $H_1$, $d$ and $\widetilde{\nu}$, where $s$ is kept fixed at
$s_{phys}$. \par

\noi Point C: the true minimum of $V_{(M+1)SSM} 
(H_1, d, \widetilde{\nu},s)$ with respect to $H_1$, d, $\widetilde \nu$ and s.
\par

First we compare the depths of the points A and B: If the effective MSSM 
(with $\bar \mu = \lambda s_{phys}$) has a "UFB" problem, point B is lower 
than point A. The analysis of the constraints on the parameters, which follow
from the condition that point B is not lower than point A, proceeds as in the
MSSM: the fact that $V_{(M+1)SSM}(H_1, d, \widetilde{\nu},s_{phys})$ contains 
an additional "constant" term $V_s(s_{phys})$ plays no role, since this term
contributes equally to the CCB minima and the physical minimum, and does hence
not affect the comparison of their respective depths. \par

Second, point C is, by construction, always deeper than point B. (The true 
minimum of $V_{(M+1)SSM}(H_1, d, \widetilde{\nu},s)$ can well be assumed for 
$s=d=0$.) It is possible, although
quite involved, to compare directly the depths of the points A and C. However,
and quite trivially, a necessary condition for point C not to be deeper than
point A is that point B has not to be deeper than point A. The MSSM like
constraints, which are required for point B not to be deeper than point A, are
thus necessary for point C not to be deeper than point A, and for the physical
minimum to be the deepest one. Accordingly the MSSM bound (\ref{2.14}) on the 
ratio $m_0/M_{1/2}$
applies also to the (M+1)SSM with universal soft terms, if one
disregards CCB minima in "UFB" directions which are deeper than the standard 
minimum (although the standard minimum would be quasi stable). \par

Possibly stronger constraints on the parameter space of the (M+1)SSM
could be obtained by comparing directly the depths of point A with point C, but
in the following we restrict ourselves to the conclusions which can be drawn
from the MSSM like analysis (the comparison of the depths of points A and B).
\par

The constraints (\ref{2.14}) on the parameter space of the
(M+1)SSM (with universal soft terms) are more dramatic than in the
MSSM: For small Yukawa couplings $\lambda$, $k$ the soft terms $A_k$,
$m_s^2$ of the (M+1)SSM are only weakly renormalized between the scales
$M_{GUT}$ and $M_Z$, and the necessary inequality (\ref{2.19}) becomes,
in terms of $A_0$ and $m_0$, 
    
\beq
\label{2.23}
|A_0| \ \gsim \ 2.9 \ m_0 \quad .
 \eeq

Clearly, the (M+1)SSM inequality (\ref{2.23}), together with
(\ref{2.14}), turns into a lower bound on the ratio $|A_0|/M_{1/2}$:

\beq
\label{2.24}
|A_0|/M_{1/2} \ \gsim \ 0.9 - 2.9 \quad .
\eeq 

On the other hand, the inequality (\ref{2.2}) provides upper bounds on
$|A_0|$ (which are somewhat weaker for $A_0$ positive; therefore the 
constraints discussed below apply actually to $|A_0|$). The remaining
region in the parameter space of the (M+1)SSM is thus quite constraint,
and can be represented in the $A_0/m_0 - m_0/M_{1/2}$ plane in Fig.~1:
Here, the vertical full and dashed lines at $m_0/M_{1/2} = 0.3$ and 1.0
respectively represent the lower 
limits (\ref{2.14}) on this ratio, which depend on $\tan \beta$ or $h_t$. 
These bounds apply to the MSSM as well as to the (M+1)SSM.  The lower limit 
(\ref{2.23}) on $A_0/m_0$ is represented by a horizontal line, and is 
specific to the (M+1)SSM. (For large Yukawa couplings $\lambda$, 
$k \ \gsim \ 10^{-2}$, where the renormalisation of $A_k$ and $m_s$ can be 
non-negligible, this lower limit can be somewhat weaker). The upper limit on 
$A_0/m_0$ from the inequality (\ref{2.2}) is represented by a full line. 
(In the MSSM this upper limit on $A_0/m_0$ is weaker and depends on $\mu$, 
cf. the remark below (\protect{\ref{2.2}})). \par

Non-vanishing vevs of $\langle T_R \rangle $ are possible (but not necessarily
deeper than the standard minimum) as soon as $m_{T_R}^2$ is negative. Within 
the remaining regions in the parameter space the value of $\rho$, which 
minimizes $m_{T_R}^2$, is $\rho \simeq 0.75$. The dotted line in Fig. 1, which
limits $m_0/M_{1/2}$ from above, corresponds to $m_{T_R}^2 = 0$
according to eq. (\ref{2.4}) with this value of $\rho$: To the right of
this line, $m_{T_R}^2$ can be negative (if $\rho$ is close to 0.75), and a
minimum with $\langle T_R \rangle \neq 0$ can even be deeper than the standard
one if, at the same time, the scale of the soft terms and hence $|m_{T_R}^2|$
is large compared to $M_Z^2$. \par

Clearly, the dotted line does not represent a strict upper bound on 
$m_0/M_{1/2}$. However, for several reasons the number of allowed points in
the parameter space decreases as we move to the right of the dotted line 
(towards larger values of $m_0/M_{1/2}$): First, for $\lambda, \kappa \lsim
10^{-2}$, as it is the case in most of the parameter space, the lower limit
on $A_0/m_0$ from eq. (2.19) is 3.0 rather than the more generous one of
2.9 indicated in eq. (2.23). Since the upper limit on $A_0/m_0$ from eq. (2.2) 
approaches also 3.0 for $m_0/M_{1/2}$ large, quite some fine-tuning is required
in this regime. Second, since here we have $M_{1/2} << m_0, |A_0|$, it becomes
increasingly difficult to satisfy the phenomenological lower limits on the
chargino and/or gluino masses. Finally, in order to keep $m_{T_R}^2$ positive 
or not too negative, one finds from eq. (2.4) that $\rho$ has to approach 1 for
$m_0 >> M_{1/2}$. Then it becomes difficult to satisfy the phenomenological
upper bound on the top quark mass. (The corresponding precise constraint on
$\rho$ is, however, quite sensitive to 2 loop contributions to the RGEs and the
effective potential.) \par

From our numerical scan of the parameter space of the (M+1)SSM employing 1 loop
RGEs and the 1 loop Coleman-Weinberg corrections to the effective potential (the
unknown threshold corrections at the GUT scale make a consistent 2 loop analysis
impossible) we find, due to the combined constraints listed above, no
phenomenologically acceptable points in the parameter space with 
$m_0/M_{1/2} \ > \ 7$. This upper limit is indicated by a dash-dotted 
line in the region of interest. The dashed region is thus the only
remaining one in the (M+1)SSM, if the standard minimum of the scalar
potential is required to be the absolute one, regardless of its decay
rate. \par

What would be the phenomenological implications of these constraints on
the parameter space of the (M+1)SSM? First, the lower limit (\ref{2.24})
on $A_0$ effects most directly the singlet sector of the model: For a
small Yukawa coupling $\lambda$ the singlet neutralino $\widetilde{S}$
(the singlino), and the singlet CP-even and CP-odd scalars are almost
pure states. The mass $M_{\tilde{S}}$ of the singlino is approximately
given by $2k\bar{s}$; after replacing $A_k$ and $m_s^2$ in the
expression (\ref{2.18}) for $\bar{s}$ by $A_0$ and $m_0^2$, the singlino
mass satisfies 

\beq
\label{2.25}
|M_{\tilde{S}}| \ \gsim \ {2 \over 3} \ |A_0| \quad . 
\eeq   

\noi From (\ref{2.24}) one then finds that the singlino can not be
lighter than the lightest neutral gaugino (typically the bino with a
mass $M_{\widetilde{B}} \sim .41 M_{1/2}$). Accordingly a ``singlino
LSP scenario'', which leads to unconventional signatures of sparticle
production in the (M+1)SSM 
\cite{19r}, would not be possible. Similarly, the masses of the CP-even
and CP-odd quasi singlet scalars are bounded from below by $|A_0|/3$ and
$|A_0|$, respectively, and would be quite heavy. \par

Second, the lower limits (\ref{2.23}) and (\ref{2.24}) on $A_0$ affect
both the renormalisation and mixings of sfermions with large Yukawa
couplings, notably the physical stop masses: Whereas the physical masses 
$m_{sq}$ of the squarks of the first generations generally satisfy

\beq
\label{2.26}
m_{sq} \ \gsim \ .9 \ M_3 
\eeq   

\noi in terms of the gluino mass $M_3$, the mass of the lightest stop
$m_{st1}$ is even bounded from above in terms of $M_3$, according to our
scan of the parameter space of the (M+1)SSM with CCB and UFB
constraints: 

\beq
\label{2.27} 
m_{st1} \ \lsim \ 80 \ \hbox{GeV} \ + .8 \ M_3 \quad .
\eeq  

Altogether, the phenomenology of the (M+1)SSM with CCB and UFB
constraints will resemble to a large extent the one of the MSSM -- with
additional, albeit quite heavy, states in the neutralino and Higgs
sectors -- and with additional constraints on the parameters and
sparticle masses. 

\mysection{Summary and Conclusions} 
\hspace*{\parindent}
The aim of the present paper is the study of constraints on the
parameter space of the (M+1)SSM with universal soft terms, which arise
from the absence of global minima of the scalar potential with vevs of
sleptons or squarks. We have considered the $D$-flat ``CCB'' and the
conventional ``UFB'' directions in field space, which are comparatively
easy to analyse. The consideration of additional ``improved'' directions
could nothing but strengthen the constraints obtained above. \par

The most important result is the fact that the MSSM ``UFB'' bounds on
the ratio $m_0/M_{1/2}$ apply also to the (M+1)SSM. This is highly
non-trivial, since the cor\-res\-pon\-ding dangerous directions in field
space are no longer $F$-flat in the (M+1)SSM. Nevertheless, this result
follows in a quite straightforward way after the minimization of the $s$
dependent part of the potential with respect to $s$, neglecting
interactions with the non-singlet sector of ${\cal O}(\lambda)$. (One
can check that the relative error of the vacuum energy induced by this
procedure is of ${\cal O}(\lambda^2)$; eventually, improved bounds in
the (M+1)SSM could be studied in the regime where $\lambda$ is not
small.) \par 

The resulting constraints on the parameter space of the (M+1)SSM are
more important than in the MSSM: Since here the soft susy breaking
terms have to satisfy the additional inequality $A_0 \ \gsim \ 2.9 \
m_0$, a lower limit on $m_0/M_{1/2}$ from ``UFB'' bounds implies a lower
limit on $A_0/M_{1/2}$, which has no analog in the MSSM. The constraints 
on the ratios of soft terms have been summarized in Fig.~1, according to
which the allowed range of the ratio $A_0/m_0$ is narrowed by the lower
``UFB'' bound on $m_0/M_{1/2}$. The fact that the ratio $A_0/m_0$ is
confined into a quite narrow region around $\sim 3$ for $m_0/M_{1/2}$
large, has already been realized in previous analyses of the parameter
space of the (M+1)SSM, where the inequality (\ref{2.2}) has been taken 
into account \citd{10r}{19r}. \par

Still, one can avoid most of these constraints -- as in the MSSM -- if
one is ready to accept the standard $SU(2) \times U(1)_Y$ symmetry
breaking vacuum as a local and not global minimum of the scalar
potential. Then, instead, constraints on the early cosmology 
-- the inflationary potential and the corresponding reheating
temperature -- can be derived from the condition to end up in the local
standard vacuum. The tunneling rates associated with the decay of the
standard vacuum have not been considered here, but they will again
correspond to the ones of the MSSM for $\lambda$ small. \par 

The phenomenological implications of the CCB and UFB constraints in
the \break \noindent (M+1)SSM are essentially 
upper bounds on the lightest stop mass as a function of the gluino mass
and, most importantly, lower bounds on the quasi singlet states, which
rule out a ``singlino LSP scenario''. Hence, if the singlino LSP
scenario could nevertheless be confirmed through its additional cascade decays
and -- possibly -- displaced vertices \cite{19r}, the metastability of our 
present vacuum would be established. \par
\vspace{1cm}

\noi  {\large{\bf{Acknowledgement}}}
\par
\noi It is a pleasure to thank Steven Abel for stimulating discussions.

\newpage

\noi{\Large \bf  Figure Caption} \par \vskip 1 truecm
\begin{description}
\item{\bf Fig. 1:} Constraints on $A_0/m_0$ (for any sign of $A_0$, cf. the
discussion below eq. (2.24)) and $m_0/M_{1/2}$ from the absence of deeper 
charge and colour breaking minima in the (M+1)SSM. The lower limits on 
$m_0/M_{1/2}$ from (\ref{2.14}) are represented by vertical full and dashed
lines. The upper and lower limits on $A_0/m_0$ from (\protect{\ref{2.2}}) and 
(\protect{\ref{2.23}}), respectively, are indicated by full lines. The upper
limit on $m_0/M_{1/2}$, which follows from $m_{T_R}^2 > 0$ with $\rho = 0.75$
from eq. (2.4) is shown as a dotted line, and the upper limit on $m_0/M_{1/2}$, 
which follows from our numerical results, by a vertical dash-dotted line. 
The dashed region is the only allowed one in the (M+1)SSM. 
\par \vskip 3 truemm
\end{description}

\newpage
\begin{figure}[p]
\unitlength1cm
\begin{picture}(1,1)
\put(-1.0,-15.0){
\epsfxsize=20cm
\epsfysize=28cm
\epsffile{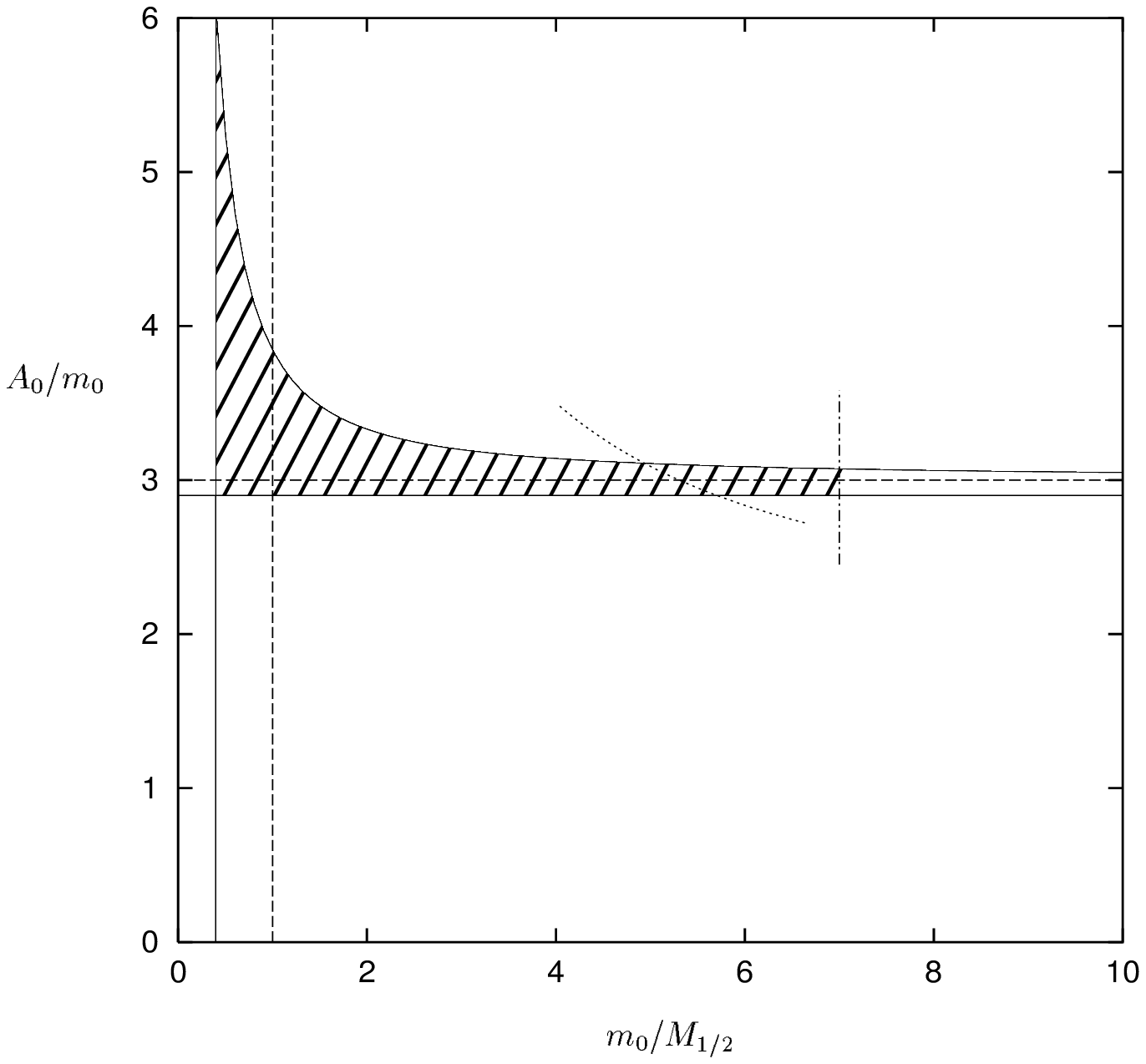}}
\put(8.0,-6) {\bf Fig. 1}
\end{picture}\par
\end{figure}

\newpage
\def\labelenumi{[\arabic{enumi}]}
\noindent
{\Large \bf References} \par \vskip 1 truecm
\ben
\item\label{1r} J. F. Fr\`ere, D. R. T. Jones, S. Raby, Nucl. Phys. {\bf
B222} (1983) 11.   
\item\label{2r} L. Alvarez-Gaum\'e, J. Polchinski, M. Wise,
Nucl. Phys. {\bf B221} (1983) 495.     
\item\label{3r} M. Claudson, L. Hall, I. Hinchliffe, Nucl. Phys. {\bf
B228} (1983) 501.   
  \item\label{4r} J.-P. Derendinger, C. A. Savoy, Nucl. Phys. {\bf B237}
(1984) 307.   
\item\label{5r} U. Ellwanger, Phys. Lett. {\bf B141} (1984) 435. 
\item\label{6r} M. Drees, M. Gl\"uck, K. Grassie, Phys. Lett. {\bf B157}
(1985) 164. 
\item\label{7r} J. Gunion, H. Haber, M. Sher, Nucl. Phys. {\bf B306}
(1988) 1.  
\item\label{8r} H. Komatsu, Phys. Lett. {\bf B215} (1988) 323.     
\item\label{9r} G. Gamberini, G. Ridolfi, F. Zwirner, Nucl. Phys. {\bf
B331} (1990) 331.  
\item\label{10r}U. Ellwanger, M. Rausch de Traubenberg, C. A. Savoy,
Phys. Lett. {\bf B315} (1993) 331, and Nucl. Phys. {\bf B492} (1997) 21.  
\item\label{11r} P. Langacker, N. Polonsky, Phys. Rev. {\bf D50} (1994) 2199.
\item\label{12r} J. A. Casas, A. Lleyda, C. Mu$\tilde{\rm n}$oz,
Nucl. Phys. {\bf B471} (1996) 3, Phys. Lett. {\bf B380} (1996) 59,
Phys. Lett. {\bf B389} (1996) 305; \\ 
H. Baer, M. Brhlik, D. Castano, Phys. Rev. {\bf D54} (1996) 6944; \\
J. A. Casas, hep-ph/9707475; \\   
S. Abel, B. Allanach, Phys. Lett. {\bf B431} (1998) 339; \\
J. A. Casas, A. Ibarra, C. Mu$\tilde{\rm n}$oz, hep-ph/9810266.
\item\label{13r} A. Riotto, E. Roulet, Phys. Lett. {\bf B377} (1996) 60; \\
I. Dasgupta, R. Rodemacher, P. Suranyi, hep-ph/9804229.    
\item\label{14r} J. A. Casas, S. Dimopoulos, Phys. Lett. {\bf B387}
(1996) 107.  
\item\label{15r} A. Kusenko, P. Langacker, G. Segre, Phys. Rev. {\bf
D54} (1996) 5824.  
\item\label{16r} S. Abel, C. A. Savoy, hep-ph/9803218.
\item\label{17r} S. Abel, C. A. Savoy, hep-ph/9809498; \\
S. Abel, T. Falk, hep-ph/9810297. 
\item\label{18r} P. Fayet, Nucl. Phys. {\bf B90} (1975) 104; \\
H. P. Nilles, M. Srednicki, D. Wyler, Phys. Lett. {\bf B120} (1983) 346; \\
J. Ellis, J. F. Gunion, H. E. Haber, L. Roszkowski, F. Zwirner,
Phys. Rev. {\bf D39} (1989) 844; \\
L. Durand, J. Lopez, Phys. Lett. {\bf B217} (1989) 463; \\
M. Drees, Int. J. Mod. Phys. {\bf A4} (1989) 3635; \\
T. Elliott, S. F. King, P. L. White, Phys. Rev. {\bf D49} (1994) 2435; \\
S. F. King, P. L. White, Phys. Rev. {\bf D52} (1995) 4183. 
\item\label{19r} U. Ellwanger, C. Hugonie, Eur. Phys. J {\bf C5} (1998)
723, and hep-ph/9812427.
\item\label{20r} S. Coleman, E. Weinberg, Phys. Rev. {\bf D7} (1973) 1888; \\
E. Gildener, Phys. Rev. {\bf D13} (1976) 1025; \\
U. Ellwanger, Nucl. Phys. {\bf B238} (1984) 665; \\
M. Sher, Phys. Rep. {\bf 179} (1989) 273.
\item\label{21r} J. A. Casas, J. K. Espinosa, M. Quiros, A. Riotto,
Nucl. Phys. {\bf B436} (1995) 3.
\item\label{22r} A. Strumia, Nucl. Phys. {\bf B482} (1996) 24; \\
T. Falk, K. A. Olive, L. Koszkowski, A. Singh, M. Srednicki,
Phys. Lett. {\bf B396} (1997) 50.
\item\label{23r} S. Abel, B. Allanach, Phys. Lett. {\bf B415} (1997) 371.

\een

\end{document}